\documentclass[12pt,superscriptaddress,preprintnumbers,amsmath,amssymb]{revtex4}
\usepackage{graphicx}% Include figure files
\usepackage{epsfig}  % Include figure files
\usepackage{epsf}    % Include figure files
\usepackage{dcolumn}% Align table columns on decimal point
\usepackage{bm}% bold math

\def\gev{{\rm GeV}}
\def\incl{B \rightarrow X_s\, l^+ \, l^-}
\def\bkll{B \rightarrow K\, l^+ \, l^-}
\def\bsg{B_s \rightarrow l^{+}\, l^{-}\, \gamma}
\begin{document}

\title{Bounds on Tensor operator contribution  to 
$\bsg$}

\author{Ashutosh Kumar Alok}
\affiliation{Indian Institute of Technology Bombay, Mumbai-400076, India and \\
Tata Institute of Fundamental Research, Homi Bhabha Road, Mumbai 400005, India}

\author{S. Uma Sankar}
\affiliation{Indian Institute of Technology Bombay, Mumbai-400076, India}

%\date{\today}

\begin{abstract} 

We consider the effect of new physics interactions in the form of
tensor operators on the branching ratio of 
$B_{s}\rightarrow l^{+}l^{-}\gamma$ where $l=e,\,\mu$. 
We find that the present data on the inclusive branching ratio
$B\rightarrow X_s l^{+}l^{-}$ puts a strong constraint on the new physics tensor couplings and does not
allow a large  enhancement in 
${\cal{BR}}(B_{s}\rightarrow l^{+}l^{-}\gamma)$
beyond its standard model predictions. Large enhancement in 
${\cal{BR}}(B_{s}\rightarrow l^{+}l^{-}\gamma)$
due to new physics in the form of vector/axial-vector, scalar/pseudoscalar
and magnetic dipole operators has already been ruled out. Thus we conclude
that no new physics can provide a large enhancement in ${\cal{BR}}(B_{s}\rightarrow l^{+}l^{-}\gamma)$ and hence it is expected to be
observed in future
experiments with a branching ratio close to its
standard model expectation.
\end{abstract}

\maketitle

\newpage
%%%%%%%%%%%%%%%%%%%%%%%%%%%%%%%%%%%%%%%%%%%%%%
\section{Introduction}
\label{intro}
%%%%%%%%%%%%%%%%%%%%%%%%%%%%%%%%%%%%%%%%%%%%%%%
The quark level interaction $b\rightarrow sl^{+}l^{-}$, where $l=e,\,\mu$, is responsible
for the (a) inclusive semi-leptonic decay $B \to X_s l^+ l^-$
(b) exclusive semi-leptonic decays
$B\rightarrow(K^{*},K)l^{+}l^{-}$, (c) purely leptonic  decays
$B_{s}\rightarrow l^{+}l^{-}$ and also (d) leptonic radiative decays
$B_{s}\rightarrow l^{+}l^{-}\gamma$. The inclusive and exclusive 
semi-leptonic decays
have been observed experimentally 
\cite{babar04_incl,belle05_incl,babar-03,babar-06,belle-03,hfag}
with branching ratios close to their Standard Model (SM) predictions
\cite{ali-02,lunghi,kruger-01,isidori}. However the decays 
$B_{s}\rightarrow l^{+}l^{-}$ and $B_{s}\rightarrow l^{+}l^{-}\gamma$ 
are yet to be observed. 

The mode $B_{s}\rightarrow l^{+}l^{-}$ is helicity suppressed and
the SM predictions for the branching ratios of the decays 
$B_{s}\rightarrow e^{+}e^{-}$ and $B_{s}\rightarrow\mu^{+}\mu^{-}$ 
are $(7.58\pm3.5)\times10^{-14}$ and $(3.2\pm1.5)\times10^{-9}$ 
respectively \cite{buras-03}. In \cite{aloksankar-05} we studied 
the relation between the semi-leptonic
decays $B\rightarrow(K^{*},K)l^{+}l^{-}$ and the purely leptonic
decays $B_{s}\rightarrow l^{+}l^{-}$. We found that if new physics
interactions are in the form of vector/axial-vector operators then
the present data on the branching ratio of 
$B\rightarrow(K^{*},K)l^{+}l^{-}$, ${\cal{BR}}(B\rightarrow(K^{*},K)l^{+}l^{-})$, doesn't
allow a large enhancement in 
${\cal{BR}}(B_{s}\rightarrow l^{+}l^{-})$
beyond the SM predictions. But if new physics interactions
are in the form of scalar/pseudoscalar operators then the present data
on ${\cal{BR}}(B\rightarrow(K^{*},K)l^{+}l^{-})$ doesn't provide any useful
constraint and a large boost is possible in 
${\cal{BR}}(B_{s}\rightarrow l^{+}l^{-})$.

The decay mode $B_{s}\rightarrow l^{+}l^{-}\gamma$
is free from helicity suppression and hence has a higher branching
ratio than the purely leptonic mode in the SM 
\cite{eilam-95,aliev-97,geng-00,Dincer:2001hu,Kruger:2002gf,Melikhov:2004mk,Chen:2006ss}.
In \cite{aloksankar-06} we studied the constraints on 
${\cal{BR}}(B_{s}\rightarrow l^{+}l^{-}\gamma)$
coming from the experimentally measured values of the branching ratios
of $B\rightarrow(K^{*},K)l^{+}l^{-}$ and $B\rightarrow K^{*}\gamma$.
New physics in the form of scalar/pseudoscalar interactions
\emph{does not} contribute to $B_{s}\rightarrow l^{+}l^{-}\gamma$ \cite{src,aloksankar-06}.
Also, new physics in form of vector/axial-vector interactions does not
provide a large enhancement in ${\cal{BR}}(B_{s}\rightarrow l^{+}l^{-}\gamma)$
as these couplings are highly constrained by the data 
on $B\rightarrow(K^{*},K)l^{+}l^{-}$.
If new physics interactions are in the form of magnetic dipole
operators then the present data on $B\rightarrow(K^{*},K)l^{+}l^{-}$ doesn't
put any useful constraint but the data on $B\rightarrow K^{*}\gamma$
puts a very strong constraint \cite{babar-04}. 
This constraint makes it impossible
to get a large boost in ${\cal{BR}}(B_{s}\rightarrow l^{+}l^{-}\gamma)$.
Therefore, new physics in the form of vector/axial-vector operators
or magnetic dipole operators can not boost the branching ratio of 
$B_{s}\rightarrow l^{+}l^{-}\gamma$ much beyond its predicted SM value \cite{aloksankar-06}.

In this paper we are interested in studying the effect of new physics
in the form of tensor operators on the branching ratio of 
$B_{s}\rightarrow l^{+}l^{-}\gamma$. We see if
new physics can enhance their branching ratio by an order of magnitude or more,
relative to the SM prediction.
These operators
can appear from the exchange of multiple gluons or spin-2 particles 
and have been studied in literature in the context of the decays 
$B\rightarrow K^{*}l^{+}l^{-}$
\cite{aliev-2001(1),aliev_2001(2),aliev-2004,cornell-05}, 
$B\rightarrow Kl^{+}l^{-}$ \cite{aliev-04,aliev-01,aliev-98,Alok:2008wp,Bobeth:2007dw},
$B\rightarrow X_s l^{+}l^{-}$ \cite{fukae-99}, 
$B_{s}\rightarrow l^{+}l^{-}\gamma$ \cite{yilmaz-03,yilmaz-04,aliev-2005} 
and $B_{s}\rightarrow\nu\overline{\nu}\gamma$ \cite{cakir-03,sirvanli-03}. 
In ref. \cite{yilmaz-03,yilmaz-04,aliev-2005}, the polarization of the
final state leptons in $B_{s}\rightarrow l^{+}l^{-}\gamma$, due to these 
tensor operators, was calculated. Here we investigate the constraints on
${\cal{BR}}(B_{s}\rightarrow l^{+}l^{-}\gamma)$ coming from the present data 
on both the exclusive semi-leptonic decay $B \rightarrow K l^+ l^-$ and
the inclusive semi-leptonic decay $B \rightarrow X_s l^+ l^-$, 
assuming new physics only in the form of tensor 
operators. 

This paper is organized as follows. In section \ref{brbsg}, 
we present the theoretical expressions 
for the branching ratio of $\bsg$ considering new physics 
in the form of  
tensor operators. 
In section \ref{cons} we study the constraints on the branching ratio of $\bsg$
coming from the measurements of ${\cal{BR}}(\bkll)$ and  ${\cal{BR}}(\incl)$.
Finally in section \ref{concl}, we present 
the conclusions. 

%%%%%%%%%%%%%%%%%%%%%%%%%%%%%%%%%%%%%%%%%%%%%%%%%%%%%%%%%%%%%%%%
\section{Effective Hamiltonian and branching ratio of $\bsg$ }
\label{brbsg}
%%%%%%%%%%%%%%%%%%%%%%%%%%%%%%%%%%%%%%%%%%%%%%%%%%%%%%%%%%%%%%%%
We consider new physics in the form of tensor operators. 
The effective Hamiltonian for the quark level transition $b \rightarrow s l^+ l^-$ can be written as
\begin{equation}
{\cal{H}}(b \rightarrow s l^+ l^-) = {\cal{H}}_{SM} + {\cal{H}}_{T} \;,
\label{ltotal}
\end{equation}
where
%\begin{widetext}
\begin{eqnarray}
{\cal{H}}_{SM} &=&  \frac{\alpha G_F}{\sqrt{2} \pi} V_{tb} V^\star_{ts}
\biggl\{ C^{\rm eff}_9  (\bar{s} \gamma_\mu P_L b)    \,
\bar{l} \gamma_\mu l +
C_{10} (\bar{s} \gamma_\mu P_L b)\,   \bar{l} \gamma_\mu \gamma_5 l 
\nonumber \\ && -
2 \frac{C^{\rm eff}_7}{q^2} m_b \, (\bar{s} i \sigma_{\mu\nu} q^\nu P_R b) \,
 \bar{l} \gamma_\mu l \biggr\}\;, 
\label{sml}\\
 {\cal{H}}_{T} &=& \frac{\alpha G_F}{\sqrt{2} \pi} V_{tb} V^\star_{ts}
\biggl\{C_T ~(\bar{s} \sigma_{\mu \nu } b) ~\bar{l}\sigma^{\mu\nu}l 
 +
 i~ C_{TE} ~\epsilon^{\mu \nu \alpha\beta} ~(\bar{s} \sigma_{\mu \nu } b)
~\bar{l} \sigma_{\alpha \beta}  l  \biggr\} \;.
\label{tenl}
\end{eqnarray}
%\end{widetext}
Here $P_{L,R} = (1 \mp \gamma_5)/2$ and $q_{\mu}$ is the sum of 4-momenta of 
$l^+$ and $l^-$. $C_T$ and $C_{TE}$ are new physics tensor
couplings. In our analysis we assume that there are no additional CP phases 
apart from the single Cabibbo-Kobayashi-Maskawa (CKM) phase. Under this assumption 
the new physics couplings $C_T$ and $C_{TE}$ are real. 

The Wilson coefficients in eq.~(\ref{sml}) will be taken to
have the following constant values: 
\begin{equation} 
C_{7}^{\rm eff} = -0.29 \; , \;  
C_{9}^{\rm eff} = +4.214 \; , \; C_{10} = -4.312\;. 
\end{equation} 

%%%%%%%%%%%%%%%%%%%%%%%%%%%%%%%%%%%%%%%%%%%%%%%%%%%%%%%%%%%%%%%%%%%%%%%%%
We now consider the process $\bsg$.
The necessary matrix element is given by,
\begin{eqnarray}
\langle \;\gamma \,(k)\left| \overline{s}\,\gamma _\mu \,(1-
\gamma _5)\,b\right| B_s(p_B)\,\,\rangle &=& \frac {e}{m_{B_s}^2}\Bigg[\in _{\mu
\nu \lambda \sigma }\varepsilon ^{*\nu }q^\lambda k^\sigma
g\,(q^2)+ \,i\,\big\{ \varepsilon ^{\mu *}(kq)
\nonumber \\ &&  -
(\varepsilon^{*}q)k^\mu \big\} \,f(q^2)\Bigg]\,,
\\
\left\langle \gamma(k)\left|\overline{s}\sigma_{\mu\nu}b\right|B_{s}(p_{B})\right\rangle &=&\frac{e}{m_{B_{s}}^{2}}\epsilon_{\mu\nu\lambda\sigma}\Bigg[G\epsilon^{\lambda *}k^{\sigma}\,
+\, H\epsilon^{\lambda *}q^{\sigma}
+\, N(\epsilon^{*} q)q^{\lambda}k^{\sigma}\Bigg]\,,
\\
\left\langle \gamma(k)\left|\overline{s}i\sigma_{\mu\nu}q^{\nu}(1+\gamma_5)b\right|B_{s}(p_{B})\right\rangle &=&
\frac{e}{m_{B_s}^2} \Bigg[
\epsilon_{\mu\alpha\beta\sigma} \, \varepsilon^{\alpha\ast} q^\beta k^\sigma
g_1(q^2)
+ i\,\Big\{\varepsilon_\mu^\ast (q k) 
\nonumber \\ && 
- (\varepsilon^\ast q) k_\mu \Big\}
f_1(q^2) \Bigg]\,,
\label{eq:sigma_munu}
\end{eqnarray}
where $\epsilon_{\mu}^{*}$ is the polarization vector of the photon and
$q$ is the momentum transfer to the dilepton.
The amplitude for the process $B_{s}\rightarrow l^{+}l^{-}\gamma$
is given by,
\begin{eqnarray}
 M &=& \frac{\alpha G_F}{4 \sqrt{2} \, \pi} V_{tb} V_{ts}^* 
\frac{e}{m_{B_s}^2} \,\Bigg[
\Big\{
A_1 \epsilon_{\mu \nu \alpha \beta} 
\varepsilon^{\nu*} q^\alpha k^\beta + 
i \, A_2 \Big( \varepsilon_\mu^\ast (k q) - 
(\varepsilon^\ast q ) k_\mu \Big) \Big\}\, \bar \ell \gamma^\mu (1-\gamma_5) \ell  \nonumber \\
&+&   \Big\{
B_1 \epsilon_{\mu \nu \alpha \beta} 
\varepsilon^{\nu *} q^\alpha k^\beta 
+ i \, B_2 \Big( \varepsilon_\mu^\ast (k q) - 
(\varepsilon^\ast q ) k_\mu \Big) \Big\}\,\bar \ell \gamma^\mu (1+\gamma_5) \ell \nonumber \\
&+& i \,\epsilon_{\mu \nu \alpha \beta}  \Big\{ G \varepsilon^{\alpha \ast} k^\beta 
+ H \varepsilon^{\alpha \ast} q^\beta + 
N (\varepsilon^\ast q ) q^\alpha k^\beta \Big\}\, 
\bar \ell \sigma^{\mu\nu}\ell \nonumber \\
&+& i \, \Big\{
G_1 (\varepsilon^{\mu \ast} k^\nu - \varepsilon^{\nu \ast} k^\mu) + 
H_1 (\varepsilon^{\mu \ast} q^\nu - \varepsilon^{\nu \ast} q^\mu) +
N_1 (\varepsilon ^\ast q ) (q^\mu k^\nu - q^\nu k^\mu) \Big\}\,\bar \ell \sigma_{\mu\nu}\ell  \Bigg]\;,
\label{tensopme}
\end{eqnarray}
with
\begin{equation}
A_1=(C_{9}^{\rm eff}-C_{10})g(q^2)-2m_bC_{7}^{\rm eff}\frac{g_1(q^2)}{q^2},\,\, A_2=(C_{9}^{\rm eff}-C_{10})f(q^2)-2m_bC_{7}^{\rm eff}\frac{f_1(q^2)}{q^2},
\end{equation}
\begin{equation}
B_1=(C_{9}^{\rm eff}+C_{10})g(q^2)-2m_bC_{7}^{\rm eff}\frac{g_1(q^2)}{q^2},\,\,B_2=(C_{9}^{\rm eff}+C_{10})f(q^2)-2m_bC_{7}^{\rm eff}\frac{f_1(q^2)}{q^2};
\end{equation}
\begin{equation}
G=4C_{T}g_{1}(q^2),\,\, N=-4C_{T}\left(\frac{f_{1}(q^2)+g_{1}(q^2)}{q^{2}}\right),\,\, H=N(qk);
\end{equation}
\begin{equation}
G_{1}=-8C_{TE}g_{1}(q^2),\,\, N_1=8C_{TE}\left(\frac{f_{1}(q^2)+g_{1}(q^2)}{q^{2}}\right),\,\, H_{1}=N_{1}(qk)\;.
\end{equation}
Here the various coefficients, $A_1, A_2$ etc, in eq.~(\ref{tensopme}) are 
expressed in terms of form factors $f(q^2)$, $g(q^2)$, $f_1(q^2)$ and $g_1(q^2)$.

It is convenient to introduce dimensionless variable $x=2E_{\gamma}/m_{B_{s}}$, where $E_{\gamma}$ is the photon energy.
This gives $q^{2}=m_{B_{s}}^{2}(1-x)$ and $qk=m_{B_{s}}^{2}x/2$. The limits of integration for $x$ are
\begin{equation} 
x_{min}=0 \; , \quad x_{max}=1-4r \; ,
\end{equation}
where $r=(m^2_l/m^2_{B_s})$.
We consider only hard
photon in the process $B_{s}\rightarrow l^{+}l^{-}\gamma$. For a
photon to be observed experimentally, its minimum energy must be greater
than $25\, \rm MeV$ which corresponds to $x\geq0.01$. Therefore we take
$x=0.01$ as the lower limit of integration.

The calculation of branching ratio gives
\begin{equation}
{\cal{BR}}(B_{s}\rightarrow l^{+}l^{-}\gamma) = {\cal{BR}}_{SM}(B_{s}\rightarrow l^{+}l^{-}\gamma) + {\cal{BR}}_{T}(B_{s}\rightarrow l^{+}l^{-}\gamma)\;,
\label{tot_brg}
\end{equation}
where
\begin{equation}
{\cal{BR}}_{SM}(B_{s}\rightarrow l^{+}l^{-}\gamma)=
\left(\frac{G_{F}^{2}\alpha^{3}m_{B_s}^{5}\tau_{B_s}}{256\pi^{4}}\right)|V_{tb} V^\star_{ts}|^2 \int \phi_{\rm SM}(x) dx\;,
\end{equation}
\begin{equation}
{\cal{BR}}_{T}(B_{s}\rightarrow l^{+}l^{-}\gamma)=
\left(\frac{G_{F}^{2}\alpha^{3}m_{B_s}^{5}\tau_{B_s}}{96\pi^{4}}\right)|V_{tb} V^\star_{ts}|^2(C_{T}^{2}+4C_{TE}^{2})\int \phi_{\rm T}(x)dx\;.
\end{equation}
The functions $\phi$'s  are defined as
\begin{eqnarray}
\phi_{\rm SM}(x) &=&  x^3\,\beta_l\,\left[(C^{{\rm eff}^2}_9+C^2_{10})\frac{1-x}{3}+(C^{{\rm eff}^2}_9-C^2_{10})r\right]\left(f^2(x)+g^2(x)\right) \nonumber \\ &&
+\,4\hat{m_b}^2\,C^{{\rm eff}^2}_7\,x^{3}\beta_l\, \left[\frac{1}{3(1-x)}+\frac{r}{(1-x)^2}\right]\left(f_1^2(x)+g_1^2(x)\right) \nonumber \\ &&
+\,4\hat{m_b}\,C^{{\rm eff}}_7\,C^{{\rm eff}}_9 \,x^{3}\beta_l\, \left[\frac{1}{3}-\frac{r}{1-x}\right]\left(f(x)f_1(x)+g(x)g_1(x)\right)\;,\\
\phi_{\rm T}(x)&=& x^{3}\,\beta_l\,\left(f^2_1(x)+g^2_1(x)\right)\;,
\end{eqnarray}
where
\begin{equation}
\beta_l=\sqrt{1-\frac{4r}{(1-x)}}\;.
\end{equation}

In eq.~(\ref{tot_brg}), the interference term is neglected because it is proportional to the lepton mass. All such terms are consistently ignored in this calculation.

The $x$ dependence of the form factors is given
by \cite{eilam-95,aliev-97},
\begin{equation}
f(x)=\frac{0.8\, {\rm GeV}/m_{B_s}}{\left[1-\frac{m^2_{B_s}(1-x)}{(6.5)^2}\right]^{2}},\,\,\,\, g(x)=\frac{1.0\, {\rm GeV}/m_{B_s}}{\left[1-\frac{m^2_{B_s}(1-x)}{(5.6)^2}\right]^{2}}\;,
\label{fg}
\end{equation}
\begin{equation}
f_{1}(x)=\frac{0.68\, {\rm GeV^2}/m^2_{B_s}}{\left[1-\frac{m^2_{B_s}(1-x)}{(5.5)^2}\right]^{2}},\,\,\,\, g_{1}(x)=\frac{3.74\, {\rm GeV^2}/m^2_{B_s}}{\left[1-\frac{m^2_{B_s}(1-x)}{(6.4)^2}\right]^{2}}.
\label{f1g1}
\end{equation}

After integration we get,
\begin{equation}
{\cal{BR}}(B_{s}\rightarrow l^{+}l^{-}\gamma) = \Big[(1.33\pm 0.40) 
 +(0.11\pm 0.03) (C_{T}^{2}+4C_{TE}^{2})\Big]\times10^{-8}\;,
\label{eq:bsgamma}
\end{equation}
where the first term is the SM prediction and the second term is the additional contribution due to new physics tensor operators.
The uncertainties quoted in eq.~(\ref{eq:bsgamma}) include a $10 \%$ uncertainty in the form factors shown in eq.~(\ref{fg}) and eq.~(\ref{f1g1}).

We see that the branching ratio depends upon the values of $C_{T}$
and $C_{TE}$. Therefore in order to get bounds on 
${\cal{BR}}(\bsg)$
we need to know the values of new physics tensor couplings. 
%%%%%%%%%%%%%%%%%%%%%%%%%%%%%%%%%%%%%%%%%%%%%%%%%%%%%%%%%%%%%%%%%%%%%%%%%%%%%%%%%%%%%%
\section{Constraints on ${\cal{BR}}(\bsg)$}
\label{cons}
%%%%%%%%%%%%%%%%%%%%%%%%%%%%%%%%%%%%%%%%%%%%%%%%%%%%%%%%%%%%%%%%%%%%%%%%%%%%
In order to obtain bound on new physics tensor couplings, we first consider the semi-leptonic process $B\rightarrow Kl^{+}l^{-}$.
The decay amplitude for $B(p_1)\to K(p_2)\,l^+(p_+)\,l^-(p_-)$ is given by 

\begin{eqnarray} 
M\,(B\rightarrow K l^{+} l^{-}) &=& \frac{\alpha G_F}{2\sqrt{2} \pi} V_{tb} V^\star_{ts}   
%\nonumber \\ &\times& 
\Bigg[\left< K(p_2) \left|\bar{s}\gamma_{\mu}b\right|B(p_1)\right> \left\{C_{9}^{\rm eff}\bar{u}(p_-)\gamma_{\mu}v(p_+)  
+C_{10}\bar{u}(p_-)\gamma_{\mu}\gamma_{5} v(p_+)\right\}  
\nonumber \\ 
& & - 2\frac{C^{\rm eff}_7}{q^2} m_b \left< K(p_2)\left|\bar{s}i\sigma_{\mu\nu}q^{\nu}b\right|B(p_1)\right> \;
\bar{u}(p_-)\gamma_{\mu}v(p_+) 
\nonumber \\ & & 
+2C_T \left< K(p_2)\left|\bar{s}\sigma_{\mu\nu}b\right|B(p_1)\right>\;\bar{u}(p_-)\sigma^{\mu\nu}v(p_+)
\nonumber \\ 
& & + 2iC_{TE}\epsilon^{\mu \nu \alpha\beta}\left< K(p_2)\left|\bar{s}\sigma_{\mu\nu}b\right|B(p_1)\right>\;
\bar{u}(p_-)\sigma_{\alpha \beta}v(p_+) \Bigg]\; ,
\label{matrix}
\end{eqnarray} 
where $q_\mu = (p_1-p_2)_\mu = (p_+ + p_-)_\mu$.
The relevant matrix elements are
\begin{eqnarray} 
\left< K(p_2) \left|\bar{s}\gamma_{\mu}b\right|B(p_1)\right> &=& 
(2p_1-q)_{\mu}f_{+}(z)+(\frac{1-k^2}{z})\, q_{\mu}[f_{0}(z)-f_{+}(z)]\;, \\
\left< K(p_1)\left|\bar{s}i\sigma_{\mu\nu}q^{\nu}b\right|B(p_1)\right> &=& \Big[(2p_1-q)_{\mu}q^2-(m_{B}^{2}-m_{K}^{2})q_{\mu}\Big]\,\frac{f_{T}(z)}{m_B+m_{K}}\;, \\
\left< K(p_2)\left|\bar{s}\sigma_{\mu\nu}b\right|B(p_1)\right> 
&=& -i\Big[(2p_1-q)_{\mu}q_{\nu}-(2p_1-q)_{\nu}q_{\mu}\Big]\,\frac{f_T}{m_B+m_{K}}\; ,
\end{eqnarray}
where $k \equiv m_K/m_B$  and $\hat{m}_b \equiv m_b/m_B$. 

The form factors $f_{+,\,0,\,T}$ can be calculated in the
light cone QCD approach. Their $z\, (=q^2/m_B^2)$ dependence 
is given by \cite{ali-00} 
\begin{eqnarray} 
f(z)=f(0)\,\exp(c_1z+c_2z^2+c_3z^3)\;, 
\end{eqnarray} 
where the parameters $f(0), c_1$, $c_2$ and $c_3$ for each form 
factor are given in Table~\ref{ff-table}.  
\begin{table}[t] 
$$ 
\begin{array}{l c c c c} 
\hline 
    & \phantom{-}f(0)  &\phantom{-}c_1  & \phantom{-}c_2  & \phantom{-}c_3\\ \hline 
	f_{+}&\phantom{-}0.319^{+0.052}_{-0.041}  &\phantom{-}1.465  &\phantom{-}0.372 
&\phantom{-}0.782\\ 
f_0 &\phantom{-}0.319^{+0.052}_{-0.041}   &\phantom{-}0.633 
&\phantom{-}-0.095  &\phantom{-}0.591\\ 
f_T &\phantom{-}0.355^{+0.016}_{-0.055} 
&\phantom{-}1.478 & \phantom{-}0.373   &\phantom{-}0.700 \\ \hline 
\end{array} 
$$ 
\caption{Form factors for the $B \to K$ transition \cite{ali-00}.
\label{ff-table}} 
\end{table} 

The calculation of branching ratio gives,
\begin{equation}
{\cal{BR}}(\bkll)={\cal{BR}}_{SM}(\bkll)+{\cal{BR}}_T(\bkll)\;,
\label{excltot}
\end{equation}
where
\begin{eqnarray}
{\cal{BR}}_{SM}(\bkll) & = & B_{0k}\,\int \Big[\psi^{3/2}\left(1-\frac{\beta_l^2}{3}\right)\,(A^2+B^2)\, +\, 4\,\hat{m}_{l}^2\,\psi^{1/2}\,B^2\,(2+2k^2-z)
\nonumber \\&&
+ \, 4\,\hat{m}_{l}^2\,\psi^{1/2}\,z\,C^2\,
+\,8\,\hat{m}_{l}^2\,\psi^{1/2}\,(1-k^2)\,BC\Big]\;dz\;, 
\\
{\cal{BR}}_T(\bkll) & = & B_{0k}\,(C_T^2+4C_{TE}^2)\,\int \frac{64\,\psi^{3/2}\,z\,f_T^2(z)}{3\,(1+k)^2}\,dz\;.
\end{eqnarray}
$A, B\, {\rm and }\, C$ are linear combinations of form-factors given by  
\begin{equation}
A  =  2C^{eff}_9\,f_{+}(z)-4C^{eff}_7\hat{m}_b \frac{f_{T}(z)}{1+k},\;\; \;\;
B  =   2C_{10}\, f_{+}(z),\;\;\;\;
C  =  2C_{10}\,\frac{1-k^2}{z}\Big[f_{0}(z)-f_{+}(z)\Big]\; ,
\end{equation}
\begin{equation}
\hat{m}_l  =  m_l/m_B,\;\; \;\;
\psi  =   1+k^{4}+z^{2}-2(k^{2}+k^{2}z+z),\;\;\;\;
\beta_l  =   \sqrt{1-\frac{4\hat{m}_{l}^2}{z}},\;\;\;\; B_{0k}=\frac{G_F^2\alpha^2\tau_B}{2^{12}\pi^5}|V_{tb}V^*_{ts}|^2m_B^5\;.
\end{equation}
The limits of integration for $z$ are
\begin{equation} 
z_{min}=4\hat{m}_{l}^2 \; , \quad z_{max}=(1-k)^2 \; .
\label{bklimit}
\end{equation}

The SM branching ratio for $\bkll$ in next-to-next leading order (NNLO) is \cite{ali-02}
\begin{equation} 
{\cal{BR}}_{SM}(\bkll)=(3.5\pm 1.2) \times 10^{-7}\;.
\end{equation}
After integration we get,
\begin{equation}
{\cal{BR}}_T(\bkll) = (0.31\pm 0.09)\times (C_{T}^{2}+4C_{TE}^{2}) \times10^{-7}\;.
\end{equation}

Thus we have
\begin{equation}
{\cal{BR}}(\bkll) = \Big[ (3.5\pm 1.2) + (0.31\pm 0.09) (C_{T}^{2}+4C_{TE}^{2}) \Big]\times 10^{-7}\;.
\label{eq:bkll}
\end{equation}
Equating the R.H.S of eq.~(\ref{eq:bkll}) with the experimental value of ${\cal{BR}}(\bkll)$ given in Table \ref{table2}, we obtain
\begin{equation}
C_T^2 + 4C_{TE}^2=2.58\pm 4.17 \; .
\end{equation}
Here all the errors are added in quadrature.  Substituting the value of $C_T^2 + 4C_{TE}^2$ in eq.~(\ref{eq:bsgamma}), we get
\begin{equation}
{\cal{BR}}(\bsg)=(1.61\pm 0.62) \times 10^{-8} \; .
\end{equation}
Therefore at $3 \sigma$, the maximum possible value of ${\cal{BR}}(\bsg)$ is $3.47 \times 10^{-8}$. In SM, the $3\sigma$ upper bound on ${\cal{BR}}(\bsg)$ is $2.58 \times 10^{-8}$. Thus we see that the additional contribution of new physics tensor operators cannot boost the branching ratio of $\bsg$ much beyond its SM prediction.   

%%%%%%%%%%%%%%%%%%%%%%%%%%%%%%%%%%%%%%%%%%%%%%%%%%%%%%%%%%%%%%%%%%%%%%%%%%%%

\begin{table}[t]
\begin{center}
\begin{tabular}{|l|}
\hline
$G_F = 1.166 \times 10^{-5} \; \gev^{-2}$ \\
$\alpha = 1.0/129.0$\\
$\alpha_s(m_b)=0.220$ \cite{beneke-99} \\
$\tau_{B_s} = 1.45 \times 10^{-12}\; s$ \\ 
$m_{B_s}=5.366 \; \gev$\\
$ m_B=5.279 \; \gev$ \\ 
$m_K= 0.497 \; \gev$\\
$m_c/m_b=0.29\;$ \cite{ali-02} \\ 
$m_b=4.80\; \gev $ \cite{ali-02} \\
$V_{tb}= 1.0 $  \\ 
$|V_{ts}|= (40.6 \pm 2.7) \times 10^{-3}$ \\ 
$\left|V_{tb}V_{ts}^*/V_{cb}\right|=0.967\pm0.009$ \cite{charles} \\ 
${\cal{BR}}(\bkll)=(4.30 \pm 0.40)\times 10^{-7}$ \cite{hfag}\\
${\cal{BR}}(\incl)_{q^2>0.04\,{\rm GeV^2}}=(4.3^{+1.3}_{-1.2})\times 10^{-6}$ \cite{hfag}\\
${\cal{BR}}(B \to X_c \ell \nu)=0.1061\pm0.0016\pm0.0006$ \cite{babar04_incl_bcl}\\ 
\hline
\end{tabular}
\caption{Numerical inputs used in our 
  analysis. Unless explicitly specified, they are taken from the 
  Review of Particle Physics~\cite{pdg}. }
\label{table2}
\end{center}
\end{table}
%%%%%%%%%%%%%%%%%%%%%%%%%%%%%%%%%%%%%%%%%%%%%%%%%%%%%%%%%%%%%%%%%%%%%%%%
We now consider the inclusive decay $B\rightarrow X_{s}l^{+}l^{-}$
and see what constraints the experimentally measured value of its
branching ratio puts on the tensor operators. 

The branching ratio of $\incl$ is given by \cite{fukae-99}
\begin{equation}
{\cal{BR}}(\incl)={\cal{BR}}_{SM}(\incl)+{\cal{BR}}_T(\incl)\;,
\label{incltot}
\end{equation}
where
\begin{eqnarray}
{\cal{BR}}_{SM}(\incl) & = & B_0\,I_{SM}\; , \\
{\cal{BR}}_T(\incl) & = & B_0\,I_T\,(C_T^2+4C_{TE}^2)\;.
\end{eqnarray}
The integrals $I_{SM}$ and $I_T$ are given by
\begin{eqnarray}
I_{SM} & = & \int dz \, \Bigg[\frac{8u(z)}{z}\left\{1-z^2+\frac{1}{3}u(z)^2\right\}C_{7}^{\rm eff}
%\nonumber \\ && 
-2\,u(z) \left\{z^2+\frac{1}{3}u(z)^2-1\right\}({C_9^{\rm eff}}^2+C_{10}^2) 
\nonumber\\ && 
-16\,u(z)\,(z-1)\,C_{9}^{\rm eff}\,C_7^{\rm eff} \Bigg]\;, \\
I_T&=&16\int dz \, u(z)\Bigg[\frac{-2}{3}u(z)^{2}-2z+2\Bigg]\;,
\end{eqnarray}
where
\begin{equation}
u(z)=(1-z)\;.
\end{equation}
Here $z \equiv q^2/m_b^2 =(p_{l^+}+p_{l^-})^2/m_b^2 =
(p_b-p_s)^2/m_b^2$. 
The limits of integration for $z$ are now
\begin{equation} 
z_{min}=4m_{l}^2/{m_b}^2 \; , \quad z_{max}=(1-\frac{m_s}{m_b})^2 \; ,
\end{equation}
as opposed to the ones given in eq.~(\ref{bklimit}) for 
the exclusive decay.
The normalization factor $B_0$ is given by
\begin{eqnarray}
B_0 = {\cal{BR}}(B\to X_c e \nu) \frac{3 \alpha^2 }{16 \pi^2 }
             \frac{ |V_{ts}^*V_{tb}|^2 }{|V_{cb}|^2 } 
            \frac{1}{f(\hat{m_c}) \kappa(\hat{m_c})}\; ,
\end{eqnarray}
where the phase space factor $f(\hat{m_c}={m_c \over m_b})$, 
and the $O(\alpha_s)$
QCD correction factor $\kappa(\hat{m_c})$ 
of $b \rightarrow c e \nu$ are given by \cite{Kim}
\begin{eqnarray}
f(\hat{m_c}) &=& 1 - 8 \hat{m_c}^2 + 8 
        \hat{m_c}^6 - \hat{m_c}^8 - 24 \hat{m_c}^4 \ln \hat{m_c} \;,  \\
\kappa(\hat{m_c}) &=& 1 - \frac{2 \alpha_s(m_b)}{3 \pi} 
           \left[(\pi^2-\frac{31}{4})(1-\hat{m_c})^2 + \frac{3}{2} \right]\; . 
\end{eqnarray}

The present world average for $B(\incl)$ is \cite{hfag}
\begin{equation}
{\cal{BR}}_{\rm exp}(\incl)_{q^2>0.04\,{\rm GeV^2}}=(4.3^{+1.3}_{-1.2})\times 10^{-6}\;.
\label{inclexp}
\end{equation}
The above branching ratios are given with a cut on the di-lepton invariant
mass, $q^2>0.04\,{\rm GeV^2}$, in order to remove virtual photon contributions
and the $\pi^{0}\rightarrow ee\gamma$ photon conversion background.
We keep the same invariant mass cut, $q^2 > 0.04$ GeV$^2$, 
in order to enable comparison with the experimental data. With this
range of $q^2$,
the SM branching ratio for $\incl$ in NNLO is \cite{ali-02}
\begin{equation}
{\cal{BR}}_{SM}(\incl)_{q^2>0.04\,{\rm GeV^2}}=(4.15\pm 0.71)\times 10^{-6}\; ,
\label{inclsm}
\end{equation}
whereas $B_0 I_T = (1.47\pm0.22)\times 10^{-6}$.
Using equations (\ref{incltot}), (\ref{inclexp}) and (\ref{inclsm}), we get
\begin{equation}
C_T^2 + 4C_{TE}^2=0.10\pm1.01 \; .
\end{equation}
Thus we see that the measurements of inclusive rate put a stronger  
constraint on the tensor operators in comparison
to the exclusive decays.

Putting above value in eq.~(\ref{eq:bsgamma}), we get
\begin{equation}
{\cal{BR}}(B_{s}\rightarrow l^{+}l^{-}\gamma)=(1.34\pm 0.41 )\times10^{-8}
\end{equation}
At $3 \sigma$, the maximum possible value of ${\cal{BR}}(\bsg)$ is $2.57 \times 10^{-8}$ which is same as the $3 \sigma$ SM upper bound. Thus we see that the effect of new physics tensor operators to the branching ratio of $\bsg$ is negligible and hence a large enhancement in ${\cal{BR}}(\bsg)$ due to such operators is ruled out. As stated earlier, new physics in the form of vector/axial-vector, scalar/pseudoscalar or magnetic dipole operators cannot provide a large boost in ${\cal{BR}}(\bsg)$, hence we conclude that a large enhancement in ${\cal{BR}}(\bsg)$ due to any kind of new physics is not possible. 

%%%%%%%%%%%%%%%%%%%%%%%%%%%%%%%%%%%%%%%%%%%%%%%%%%%%%%%%%%%%%%
\section{Conclusions} 
\label{concl}
%%%%%%%%%%%%%%%%%%%%%%%%%%%%%%%%%%%%%%%%%%%%%%%%%%%%%%%%%%%%%%%%
We are interested in finding the contribution of new 
physics interactions in the form of tensor operators 
to the branching ratio of $B_s \rightarrow l^+l^-\gamma$ 
subject to the constraints coming from the measurements on
on the exclusive decay $B \rightarrow K l^+ l^-$ and the 
inclusive decay $B \rightarrow X_s l^+ l^-$. 
These operators do not contribute to purely leptonic decays. They contribute 
only to semi-leptonic and radiative leptonic decays. 
We found that the present data on $B\rightarrow X_s l^+l^-$
 put  strong constraints on these couplings and rules out large enhancement in the branching ratio of 
$B_{s}\rightarrow l^{+}l^{-}\gamma$ in comparison to the SM predictions.
In \cite{aloksankar-06} it was shown that new physics in the form of 
vector/axial vector, scalar/pseudoscalar and magnetic dipole operators cannot 
provide large enhancement in ${\cal{BR}}(\bsg)$. 
Therefore we conclude that no new physics can provide a large enhancement in the
branching ratio of $\bsg$ above its SM predictions.

\end{document}